\documentclass[twocolumn,showpacs,showkeys,preprintnumbers,amsmath,amssymb]{revtex4}

\usepackage{graphicx}
\usepackage{dcolumn}
\usepackage{bm}

\def\beq{\begin{equation}}
\def\eeq{\end{equation}}

\def\phi{{\varphi}}
\def\equal{\buildrel {\rm def} \over {=} }
\def\={ \equal } 
\def\FPU{ {Fermi--Pasta--Ulam}\ }
\def\Hfpu{ {H_f} }

\begin{document}

\preprint{APS/123-Cond Mat}

\title{Metastable states in the FPU system}

\author{Andrea Carati}
 \email{carati@mat.unimi.it}
\author{Luigi Galgani}
 \email{galgani@mat.unimi.it}
\affiliation{
    Department of Mathematics, University  of  Milano\\ 
    Via Saldini 50, 20133 Milano, Italy
}

\date{\today}

\begin{abstract}
In this letter we report numerical results giving, as a function of
time, the energy fluctuation of a \FPU system in dynamical contact with a
heat bath, the initial data of the FPU system being extracted from a
Gibbs distribution at the same temperature of the bath. The aim is to 
get information on the specific heat of the
FPU system in the spirit of the fluctuation--dissipation theorem. While
the standard equilibrium result is recovered at high temperatures, 
there exists a
critical temperature below which the energy fluctuation as a
function of time tends to an asymptotic value sensibly lower than the
one expected at equilibrium. This fact appears to  exhibit the existence
of a metastable state for generic initial conditions.  An analogous
phenomenon of metastability was up to now observed in FPU systems only
for exceptional initial data having vanishing Gibbs measure, namely 
excitations of a few low--frequency modes (as in the original FPU work).
\end{abstract}

\pacs{05.45.-a, 05.20.-y}
\keywords{FPU, specific heat, metastability}
\maketitle

The \FPU system, i.e. a chain of particles interacting through 
(nearest--neighbour) 
weakly non--linear forces, was introduced (see \cite{FPU})
with the aim of clarifying a fundamental problem in statistical
mechanics, namely to understand how quickly is equilibrium attained
(the so--called ``\emph{rate of thermalization}''). The state of the
art can be summarized as follows.
By numerical simulations with excitations of a few  low--frequency
modes, it was found that  equipartition of the
(time--averaged) 
mode energies, which is predicted by equilibrium statistical
mechanics, is attained rather quickly if the initial energy is
above a certain threshold.
Below such a threshold one finds instead (see \cite{antonio}), 
a quick relaxation to a state in which equipartition of energy obtains 
only within a packet 
of low--frequency  modes, with an
exponential tail towards the high--frequencies.
The time needed for the formation of such a packet is found to
increase as an inverse power of the specific energy as the latter is diminished. 
The subsequent evolution to the final
equilibrium state (with full equipartition) is expected  to occur on a 
much longer time scale, displaying  a totally different dependence, possibly
of exponential type, on inverse specific energy.
Semi--analytical and numerical indications were first given in
\cite{parisi1}, and later confirmations were given for example in \cite{simone}.
So, at low enough temperatures the system behaves as if
it had attained  a final equilibrium, although it is actually being in a  kind of
metaequilibrium state, somehow analogous to the familiar ones
of  glasses. 

Now, the set of initial data with excitations of a few
low--frequency modes, which are the ones dealt with in the
above--mentioned results,   is statistically irrelevant in the thermodynamic
limit, being exceptional with respect to the Gibbs measure. 
So there naturally arises the question whether an analogous phenomenon of
metaequilibrium occurs also for typical initial data, 
so as to be relevant for the foundations of Statistical Mechanics.
\begin{figure}[ht]
 \begin{center}
   \includegraphics[width=8.cm]{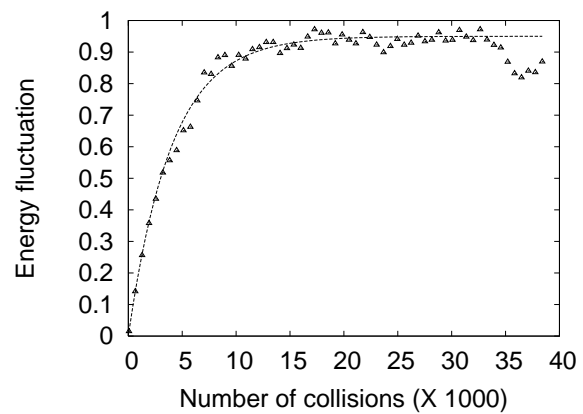}
   \caption{ \label{fig1} Energy fluctuation versus number of
   collisions (in units of a thousand) at temperature $T=1$ for 100 particles. The
   fitting curve is discussed later in the text.        }
 \end{center}
\end{figure}
On the other hand, typical initial data imply by definition  equipartition
of energy, so that it not clear which quantity should be observed in
order to test whether a 
metaequilibrium  state rather than an equilibrium one is  being
attained.
In the present paper  we propose a concrete quantity which is suitable
to this end. Moreover, we report the results of some numerical
computations, which appear to indicate that metaequilibrium states 
actually occur in a FPU system  for typical initial data at low enough
temperatures.

To this end,
we make reference to the formula
relating equilibrium specific heat and temporal energy fluctuations
according to the fluctuation--dissipation theorem, namely 
\begin{equation}\label{eq1}
  \lim_{t\to+\infty} <(E(t)-E(0))^2>_{T}/(2NT^2) = C_V(T) \ .
\end{equation}
Here one considers a system in dynamical contact with a   
heat bath at a given  temperature $T$, and $E(t)$ is the system's
energy at time $t$; furthermore,
 $<\cdot>_{T}$ denotes Gibbs average over the initial data
at the same temperature $T$, while $C_V(T)$ is the corresponding
equilibrium (canonical) specific heat (with the Boltzmann constant $k_B=1$).

The quantity that we propose as a suitable observable is then nothing
but the function $F(t,T)\=<(E(t)-E(0))^2>_{T}/(2NT^2)$ itself.  
One should compute it for an FPU system 
in dynamical contact with a heat bath at temperature $T$,  and plot it versus
time. Indeed, on the one hand, 
such a quantity by definition makes reference to typical  initial
data. On the other hand, when plotted as  a function of time, 
it starts up  from zero and should
attain, after a certain relaxation time, the equilibrium value
$C_V(T)$. The relaxation time can then be concretely estimated as a function of
tenperature $T$.  Now, one might agree that a metaequilibrium state
has been exhibited if,  for a given temperature $T$,  the  quantity  
$F(t,T)$ as a function of time is found to have
relaxed  to some definite value, sensibly lower than the equilibrium
one. Indeed, in such a case, that value should legitimately be
considered as estimating the specific heat which is observed in an actual
measurement  of time--length $t$.

The FPU system we consider is a chain of $N+2$ particles of equal mass $m$
with fixed ends, interacting through
a first neighbour quartic potential
(the so called $\beta$--model); the Hamiltonian is thus (the lower
index ``$f$'' standing for ``free'', as opposed to ``$tot$'' used
below for ``total'') 
\begin{equation}\label{eq2}
 \Hfpu (p,q)=\sum_{i=1}^{N}  \frac {p_i^2}{2m} + 
\sum_{i=0}^{N} \big[ \frac {\Omega^2}2 (q_i - q_{i+1})^2
+ \frac\beta 4 (q_i - q_{i+1})^4 \big] \ , \nonumber
\end{equation}
with  boundary conditions $ q_0=q_{N+1}=0$, while 
$\Omega$ and $\beta$ are suitable parameters.

The interaction with a heat bath is modeled as follows. In principle,
we think
of the heat bath as a perfect gas constituted by  a very large number of
particles, each of which  
interacts, via a suitable rapidly decreasing potential, with the leftmost
moving particle of the FPU system.  Each collision then produces a certain
exchange of energy between the body and the heat bath. However, it
would be practically impossible to perform the
numerical integrations for  the resulting very large number of equations of
motion.  Thus, we choose as a model the related one in which the
collisions with all the molecules of a large bath are replaced by
the successive collisions with a  \emph{single} gas particle,
the initial data of the particle before each collision
being extracted from  a Maxwellian distribution at temperature $T$;
further details are given below.
Denoting by $x,p_x$ the position and the momentum of the incoming
gas particle, we thus consider the totat system with Hamiltonian
$$
H_{tot}(p_x,x,p,q)=p_x^2/2m + V(x-q_1) + \Hfpu(p,q)\ .
$$
The interaction potential $V$ is taken of the form $V(r)=V_0\cdot
(L/r)\, \exp(-r/L)$, where $V_0$ and $L$ are the strength 
and the range respectively of the potential.  
The quantities $m$, $V_0$ and $L$ are chosen as units of mass, energy
and length respectively, 
while the parameters $\Omega$ and $\beta$ are determined by a power expansion
about the equilibrium point of a Lennard-Jones potential having 
$V_0$ as depth and $L$ as range. In this way,  the 
unusual values $\Omega=20$ and $\beta=3600$ are obtained.
\begin{figure}[ht]
 \begin{center}
   \includegraphics[width=8.cm]{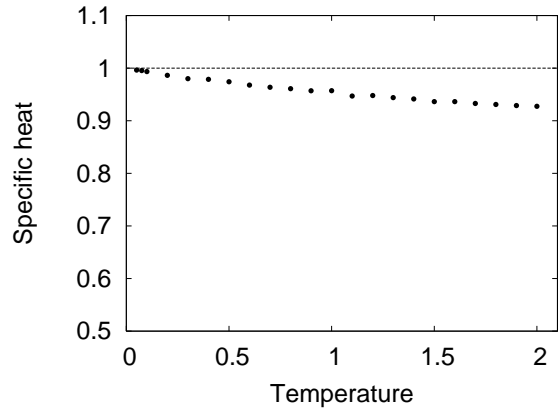}
   \caption{ \label{fig2} Specific heat versus temperature, with
   the anharmonicity taken into account. The corresponding harmonic value
   1 is also indicated.}
 \end{center}
\end{figure}

Concerning the initial data of the gas particle, for the position we take
$x_o=20\, L$ (so that the initial interaction potential is negligible), 
while the momentum $p_{xo}$ is extracted at random according to a Maxwellian
distribution at a given temperature $T$. More precisely, it  is well
known  that  the constraint on the initial position requires  a
correction factor for the Maxwellian: in fact,  the velocity has to be 
extracted from a distribution with density $ \rho(p_x)= C p_x
\exp(- p_x^2/2mT) $,  with a  normalization  constant $C$. 
Concerning the initial data for the FPU system, they  are 
extracted  from a Gibbs distribution at temperature $T$, with
reference to the full anharmonic Hamiltonian $H_{f}$ involving the
quartic terms. The technical way in which this was implemented is
described below.

Having thus chosen the initial data in the said way,  
the equations of motion corresponding to the total
Hamiltonian $H_{tot}$ are integrated by the leap--frog method, and the collision is
considered to have terminated  when the position of the gas particle becomes
again equal to  $x_o$.
At such a moment we read the value $E$ of the energy  of the FPU system.
We iterate the procedure by extracting each time a new random velocity
for the incoming particle, while the data for the FPU system are left
unchanged,
i.e. the initial FPU data for the new collision are just equal to the final ones
of the previous collision.
Thus one obtains a sequence $\{E_n\}$ of energy values, and
the exchanged energy up to ``time'' $n$
is then $(\Delta E)_n\= E_n -E_0$, where $E_0$ is the initial energy.
\begin{figure*}[th] 
 \begin{center}
  \includegraphics[width=17.cm]{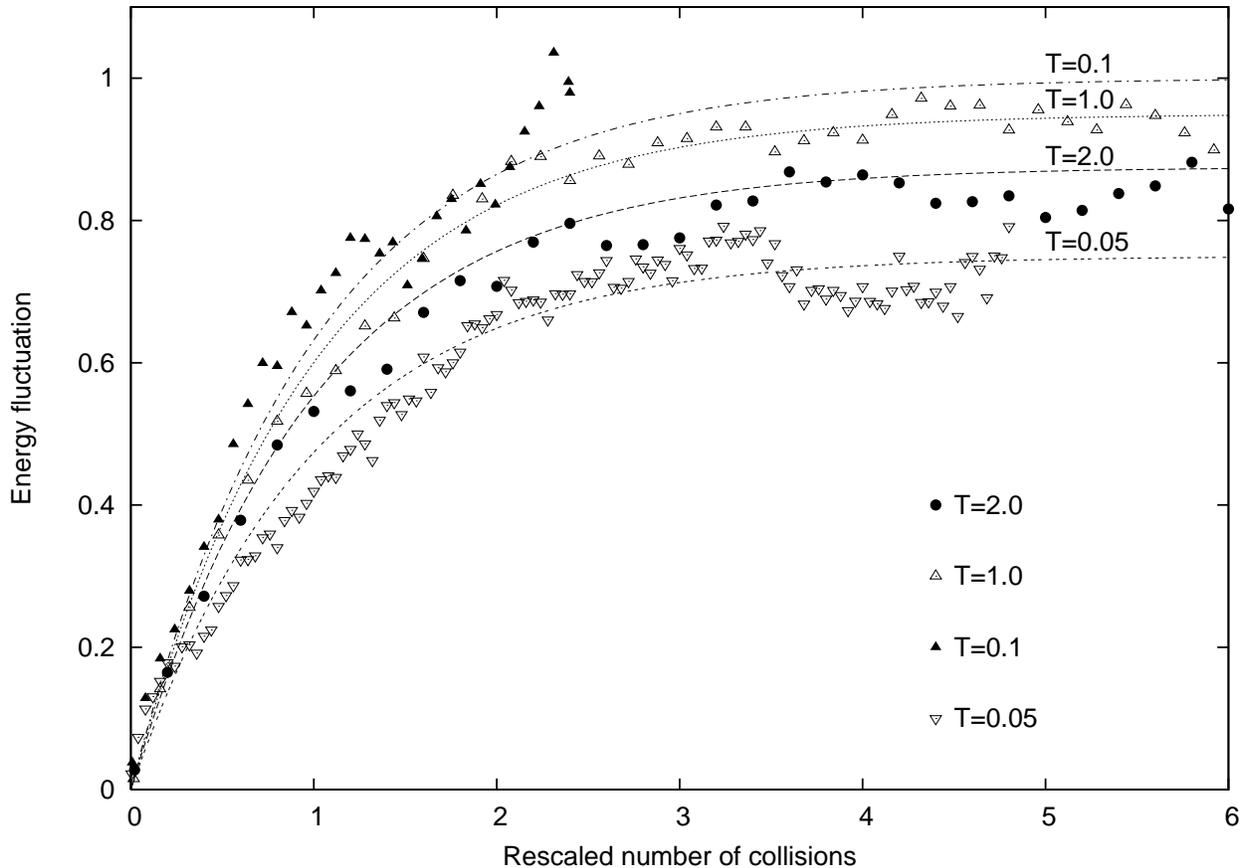}
  \caption{\label{fig3} Energy fluctuation versus rescaled number of
  collisions, with the corresponding fits. The simulation for $T=0.05$
  took a one month work of  a parallel machine using 52 CPUs.}
 \end{center}
\end{figure*}
In order to estimate  the quantity $<(E_n-E_0)^2>_{T}$ which appears in
(\ref{eq1}), one has to repeat the above procedure a sufficient 
number $K$ of times, each time choosing the initial data for the FPU
system at random from a Gibbs distribution at temperature
$T$. Thus, one has $K$ initial energies $E_0^i$, $i=1,...,K$,
and $K$ energy sequences $\{E_n^i\}$, each one corresponding to  $n$
collisions, and finally one sets 
$ <(E_n-E_0)^2>_{T} =  \sum_{i=1}^{K} (E_n^i -E_0^i)^2/K$.

An example of the results obtained in such a way is illustrated in 
Figure \ref{fig1}, where the quantity
$<(E_n-E_0)^2>_{T}/2NT^2$ , which from now on will be called the
``\emph{energy fluctuation}'' ( but might rather be called mean squared
``\emph{energy jump}'') is plotted versus  the number $n$ of collisions.
The figure corresponds  to the case of a FPU system
with $N=100$ moving  particles, at  temperature $T=1$ (in our units),
with  $K=640$. One sees that the energy fluctuation
starts up growing from $0$, and after a
number $n\simeq 3\,\cdot  10^4$  of collisions appears to have
attained  an asymptotic
value $\simeq 0.95$. We recall that, for a purely
quadratic FPU Hamiltonian,  the equilibrium value of $C_V(T)$ is
independent of temperature and equal to 1.
To decide whether the observed ``final value'' agrees with the
predictions of equilibrium statistical mechanics,
the canonical value of $C_V$ at $T=1$ corresponding to the full anharmonic
FPU system $H_f$ has to be evaluated. 

In order to illustrate how this was performed, 
we start describing  how  the anharmonic contribution was taken into
account in the related problem of extracting the initial data
 according to the Gibbs distribution proportional to
$\exp(-\Hfpu(p,q)/T)$. To this end we make  use of the so called  
``\emph{rejection method} \cite{prob}.  Namely, we start considering
only the quadratic part (say $H_2(p,q)$) of the FPU Hamiltonian $\Hfpu(p,q)$, 
because in such a case 
the normal modes are  distributed as independent gaussian
variables. Having 
extracted the values for the normal modes according to the harmonic
Hamiltonian,  we then read the numbers 
$E_0=\Hfpu(p,q)$ and $\tilde E_0=H_2(p,q)$.  
The values are rejected if, having chosen at random a
value $z$ in $(0,1)$, one has $z > \exp((\tilde E_0-E_0)/T)$, while
they are accepted in the opposite case.
The sequence of accepted values is known to be  distributed according
to a density proportional to $\exp(-\Hfpu(p,q)/T)$.
In this way we are also able to numerically evaluate the canonical
specific heat for the full Hamiltonian $H_f$, by just
evaluating at each temperature $T$ the standard deviation of the
distribution and using the familiar canonical formula relating
standard deviation and specific heat. 

The results of such a procedure are reported in Figure~\ref{fig2}, where
the equilibrium (canonical) specific heat $C_V$ for the full Hamiltonian $H_f$ is
plotted versus temperature $T$. The computations were performed 
by extracting a sample of  $K=10^6$ data at
each temperature, and using the formula 
$NT^2C_V= {\sum_{i=1}^{K}(E_0^i)^2}/{K}-\left(\sum_{i=1}^{K}E_0^i/{K}\right)^2$.
From Figure~\ref{fig2}, one sees  that at temperature $T=1$ one has
for the equilibrium specific heat a value $\simeq 0.95$, which is in
very good agreement with the asymptotic value of the dynamical energy
fluctuation reported in Figure~\ref{fig1}.
We conclude that at temperature $T=1$ relation (\ref{eq1}) 
is well satisfied and that
the corresponding relaxation time is shorter than the 
available computation time.

Now one can proceed to compute the energy fluctuation as a function of
time for other values of temperature $T$, in order to check whether 
an asymptotic values is actually attained and whether the latter 
agrees  with the final one expected according to (\ref{eq1}).  
The first result we find is that the number of collisions needed
to attain an asymptotic regime  increases  substantially as  temperature is
lowered. 
This is expected, since the mean velocity  of the gas particle
decreases, and this leads to smaller energy exchanges (see for example
\cite{levy}  for the case of diatomic molecules).
The increase of the relaxation time actually turns out to be so steep, that 
we cannot afford to observe a relaxation at all in simulations with $T
< 0.05$,  and even for $T=0.05$ 
it took a month to observe a relaxation with a computer using
52 parallel processors.  

From several simulations at $T>0.05$  we found that the number of 
collisions needed to observe a relaxation 
scales as $1/T^2$. This  was  obtained
by   fitting  the family of curves
$F(n,T)\=<(E_n-E_0)^2>_{T}/(2NT^2)$ to  curves of the form  
$F(n,T)=A_{\infty}(T)(1 -
\exp(-n/n_0(T))$, which gave  $n_0(T)=4\cdot10^3/T^2$.

Our main result is reported in Figure~\ref{fig3}, where
 the energy fluctuation (as defined above) is plotted versus 
the rescaled number of collisions $n/n_{0}(T)$, for
different temperatures; the above mentioned interpolations are also
 reported.  First we considered a temperature larger than $T=1$,
 i.e. $T=2$. The expected asymptotic value is now smaller, namely
 $\simeq 0.92$, due to the increased anharmonic contribution (which
 has a negative sign) to the specific heat. The energy
 fluctuation appears indeed to have relaxed to a smaller value than at
 $T=1$, and  actually the best fit gives $A_{\infty}=0.87$, so that there is
 at least a qualitative agreement. The
 relevant step however is now to go to lower temperatures, where
 qualitative discrepancies with respect to equilibrium statistical
 mechanics may occur. Notice by the way that the anharmonic
 contribution to the specific heat steaily decreases with decreasing
 temperature, so that the equilibrium value steadily approaches the
 harmonic value 1, and is essentially indistinguishable from it for
 example at $T=0.1$. This trend of approaching the harmonic
 equilibrium value is followed also by the dynamical asymptotic value
at $T=0.1$, as the results of Figure~\ref{fig3} show. One will notice
 that the computations were performed in such a case only for a rather
 short time. The
 reason was that the computer times become formidable with decreasing
 temperatures. So we decided that the indications available at $T=0.1$
 were sufficient, and concentrated our attention to the case $T=0.05$,
 with a run that, as mentioned above, took one month of
 computation. The results were however quite rewarding. Indeed, here
 too a rather good approach to some asymptotic value is obtained. But
 there is however  an inversion in the trend of the asymptotic
 value because  the present one, instead of being still nearer to 1,
 turns out to have sensibly diminished,  being $\simeq 0.75$.

We interpret this result as a strong indication that  a metastable state has
been attained.
Indeed here too, in analogy with the case of initial excitations of a few
low--frequency modes, we may expect that  
the energy fluctuation will
eventually attain the equilibrium canonical limit.  
But the results of Figure~\ref{fig3} suggest that at $T=0.05$ the final
relaxation would require 
a number of collisions much larger than $n_0(T)$. In other terms, 
the time--scale to the final equilibrium is expected to be much larger
than the one leading to the relaxation observed here.
Such an occurring of (at least) two different
time--scales indeed is a characteristic feature 
of metastability  phenomena. 
Our numerical results thus appear to be a direct indication that,
below a certain critical temperature ($T=0.05$ in our units), two
time--scales, and thus a metastability phenomenon, do show up
for initial data of full  measure, in connection with measurements of
the specific heat. For a previous indication see \cite{vetri}.
It would now be interesting to explore the region of lower
temperatures. The situation is however rather hard, because
unfortunately, as typical of all metastability phenomena, 
the critical temperature also constitute an actual
bound for possible numerical experiments.

\end{document}